\documentclass[prd,aps,superscriptaddress,preprintnumbers,notitlepage,nofootinbib]{revtex4-2}
\usepackage{amssymb,amsthm,amsmath}
\usepackage{textcomp}
\usepackage{color}      
\usepackage{slashed}    
\usepackage{verbatim}
\usepackage[normalem]{ulem} 
\usepackage{hyperref}
\usepackage{soul}

\usepackage{booktabs}
\usepackage{rotating}   
\usepackage{multirow}   
 \usepackage{simpler-wick}
 
 \usepackage{subcaption}
 \usepackage{epsfig}
\begin{document}
 	
 	\title{Large CP violation  in   charmed baryon decays}
 	
 	\author {
 		Xiao-Gang He}\email{hexg@sjtu.edu.cn}
 	\affiliation{State Key Laboratory of Dark Matter Physics, 
 		Tsung-Dao Lee Institute and School of Physics and Astronomy,
 		Shanghai Jiao Tong University, Shanghai 200240, China} 
 	\affiliation{Key Laboratory for Particle Astrophysics and Cosmology (MOE) and Shanghai Key Laboratory for Particle Physics and Cosmology, Tsung-Dao Lee Institute and School of Physics and Astronomy,
 		Shanghai Jiao Tong University, Shanghai 200240, China}
 	\author {
 		Chia-Wei Liu}\email{chiaweiliu@sjtu.edu.cn	}
 	\affiliation{State Key Laboratory of Dark Matter Physics, Tsung-Dao Lee Institute and School of Physics and Astronomy,
 		Shanghai Jiao Tong University, Shanghai 200240, China} 
 	\affiliation{Key Laboratory for Particle Astrophysics and Cosmology (MOE) and Shanghai Key Laboratory for Particle Physics and Cosmology, Tsung-Dao Lee Institute and School of Physics and Astronomy,
 		Shanghai Jiao Tong University, Shanghai 200240, China}
 	\date{\today}

 	\begin{abstract}
 		
This work presents the first detailed numerical predictions of CP violation in antitriplet charmed baryon decays. Adopting the flavor $SU(3)_F$  symmetry and the final state re-scattering framework, we relate the CKM-suppressed amplitudes to the leading ones, which are proportional to $\lambda_b$ and $\lambda_d$, respectively, with $\lambda_q = V_{uq}V^*_{cq}$. Larger-than-expected CP violations are found,	reaching the order of 	$10^{-3}$, similar in magnitude to those in $D^0 \to \pi^+\pi^- ,K^+K^-$. 
This opens a promising avenue for observing CP violation in charmed baryon decays and testing the Standard Model.
 	\end{abstract}
 
 \pagenumbering{roman}
 \thispagestyle{empty}
 \begin{figure}[!h]
 	\includegraphics[width=.81\linewidth]{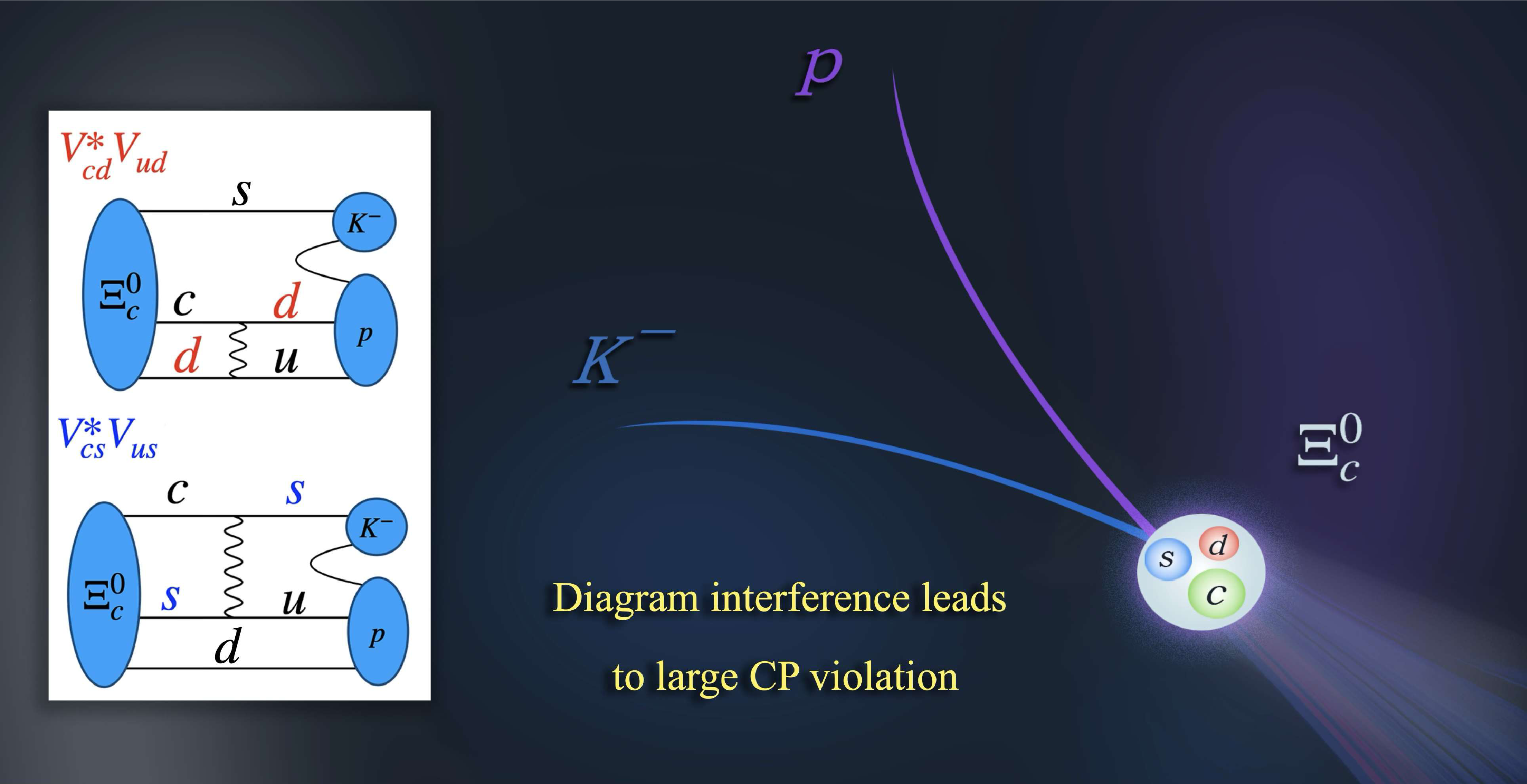}
 	\caption*{{\bf Graphic abstract:} 
Large CP violation in charmed baryon decays arises from the interference between the two contributing diagrams
 	}
 	\label{fig:gra}
 \end{figure}

  \newpage 
  
  \pagenumbering{arabic}
 	\maketitle
 
\begin{center}
 \noindent
  {\it Received: 13-Mar-2025. 
    Revised:  22-Apr-2025.
     Accepted:   27-May-2025.} 
\\[2pt]
\end{center}

 \section{
Introduction 
}
The study of CP violation is fundamentally important as it is closely related to the baryon asymmetry of our universe~\cite{Sakharov:1967dj}.  Since the CP violation from the Cabibbo-Kobayashi-Maskawa (CKM)   matrix   $V_{qq'}$~\cite{Kobayashi:1973fv} is too small to explain the observed baryon asymmetry, physics beyond the Standard Model (SM) is well motivated. Where the new CP-violating source comes from is still unknown, and the SM should be tested in all possible ways. Systems containing a charm quark  host some  interesting surprises, such as the CP-violating rate asymmetries of $D^0 \to \pi^+\pi^-,\; K^+ K^-$ observed at LHCb~\cite{LHCb:2019hro,LHCb:2022lry}, which are an order  of magnitude larger than short-distance (SD)  expectations~\cite{Lenz:2020awd}. Although there are speculations taking this as a signal for new physics beyond the SM~\cite{Schacht:2022kuj,Bause:2022jes}, it has also been argued that if long-distance (LD) contributions are properly taken into account, the SM can  
 accommodate 
 the observed sizes of  CP asymmetries~\cite{Cheng:2019ggx,Bediaga:2022sxw,Pich:2023kim}.
To reach a robust conclusion, a lattice QCD simulation is necessary.

 We propose to further  test CP violation in the charm sector using antitriplet charmed baryon $({\bf B}_c)$ decays into a low-lying octet baryon $({\bf B})$ and a pseudoscalar meson $(P)$.
The decays of charmed mesons and baryons may exhibit similar behavior. The possibility of significant CP violation 
in  \(\mathbf{B}_c\)  decays should be examined both theoretically and experimentally.
Recently, the first evidence of CP violation in baryon decays has been observed in  a heavy  $b$-baryon system~\cite{LHCb:2025ray},    with magnitudes consistent with SM estimates.   It is  of great interest to examine whether another  heavy baryon sector, the charmed baryon, also exhibits CP violation.  Our analysis, based on flavor $SU(3)_F$ symmetry, suggests that testing CP violation in this sector is feasible, potentially with enhanced effects, as seen in $D^0 \to \pi^+\pi^-, K^+ K^-$ decays. This highlights the relevance of two-body charmed baryon decays.

The amplitudes of ${\bf B}_c \to {\bf B}\,P$ are   parameterized as
$
{\cal M} = 
 	 		i \overline{u} \left(
 	F - G \gamma_5
 	\right)u_c\,, $
 where \( F  \) and \( G \) are related to the S wave and P wave, respectively, and $u_{(c)}$ is the Dirac spinor of ${\bf B}_{(c)}$. 
 If  CP were conserved, $F$ would 
 flip sign
 and  $G$ would remain invariant under the CP transformation.
One can construct the
CP-violating 
 rate asymmetry $A_{CP}
 = ({\cal B} - \overline{{\cal B}}) / ({\cal B} +  \overline{{\cal B}} 
) $
 and polarization asymmetry $A^\alpha_{CP}
 =(\alpha +  \overline{\alpha} )/2 
 $, where  ${\cal B} (\bar {\cal B})$ and $\alpha(\bar \alpha)$ represent the branching fraction and the polarization parameter for the particle (antiparticle), respectively~\cite{Lee:1957qs}.

For non-vanishing   $A_{CP}$ and $A_{CP}^\alpha$,  at least two different  weak CP-violating phases accompanied by different strong phases are required.  Without $K^0-\overline{K}^0$ mixing it is not possible for Cabibbo-favored~(CF) or doubly Cabibbo-suppressed~(DCS) processes, but possible for the singly Cabibbo-suppressed~(SCS) 
processes with two weak phases
  $\lambda_d$ and  $\lambda_s= -\lambda_d -\lambda_b$
  with $\lambda_q= V_{cq}^* V_{uq}$.  
 It is significant that the BESIII collaboration observed a sizable phase shift in $\Lambda_c^+ \rightarrow \Xi^0 K^+$ decays~\cite{BESIII:2023wrw, Wang:2024wrm}, suggesting large strong phases within the SM---a key ingredient in the study of CP violation.

Theoretical calculations, starting from the quark-level effective interaction, are challenging due to the strong interactions at the charm scale.  
No reliable results from first-principles calculations have been established.
Methods based on flavor $SU(3)_F$ symmetry considerations can simplify these calculations in a model-independent manner. 
It has been demonstrated that $SU(3)_F$ predictions fit data well~\cite{Xing:2023dni,Zhong:2022exp,Huang:2021aqu,Asymmetries,Zhong:2024qqs-a,Zhong:2024qqs-b,Yan:1992gz,He:2018joe,Geng:2023pkr}. 
Nevertheless, this approach cannot provide dynamical details.

We propose a new method for performing the $SU(3)_F$ analysis.  At the beginning, we will adopt the usual framework of the $SU(3)_F$ fit. We then reinterpret the fitted parameters in a dynamical model. In particular, we will demonstrate  how the fitted amplitudes are related to the  dynamical quantities in the framework of the final state re-scattering~(FSR). This reinterpretation  helps us   understand  the meanings of   the $SU(3)_F$ numerical analysis and give us ways to predict CP violation as will be shown later.

\section{
Materials and methods
}

Under the $SU(3)_F$,
the    effective Lagrangian  transforms as 
${{{\bf 15}}}$, $\overline{\bf 6}$ and ${\bf 3}$. The explicit tensor representations  responsible for the CF and DCS can be found in Ref.~\cite{Huang:2021aqu}.   For 
the tree-level operators of 
SCS,   we have~\cite{Buchalla:1995vs}
	\begin{equation}   
		\label{LSCS}
		{\cal L}^{\text{SCS}}_{eff} =   -\frac{G_F}{\sqrt{2}}
	 \sum_{\lambda =\pm} 
	  C_\lambda ({\cal H}_\lambda  ) ^{ij}_k  	\overline{q}_i 
		\gamma^\mu(1-\gamma_5)q^k \overline{q}_j  \gamma_\mu(1-\gamma_5)
		c  \,,
\end{equation}
where 
$G_F$ is the Fermi constant and  $C_\pm$  are the Wilson coefficients.
The  
$SU(3)_F$ 
 tensors  
are defined by 
 \begin{eqnarray}\label{Op}
({\cal H}_+)^{ij}_k &=&
	\frac{\lambda_s -\lambda_d}{2 }
	{\cal H}({\bf 15}^{s-d} )_{k}^{ij}   + 
\lambda_b 	\left( 
	{\cal H}({\bf 15}^{b} )_{k}^{ij} + 
	{\cal H}({\bf 3}_+)^{i} \delta ^{j}_k +    {\cal H}({\bf 3}_+ )^{j } \delta ^{i}_k \right)	,   
\end{eqnarray}
and 
 \begin{eqnarray}\label{Om}
({\cal H}_-)^{ij}_k &=& 		\frac{\lambda_s -\lambda_d}{2 }
	{\cal H}(\overline{{\bf 6}})_{kl}\epsilon^{lij} +  2 \lambda_b \left(   {\cal H}({\bf 3}_-)^{i} \delta ^{j}_k -    {\cal H}({\bf 3}_-)^{j } \delta ^{i}_k 
	\right) \,.
\end{eqnarray}
By construction ${\cal H}_+$ and ${\cal H}_-$ are 
symmetric and antisymmetric  in the two upper indices. 
The nonzero matrix elements
are  given as 
${\cal H}(\overline{{\bf 6}}     )_{23} = 
 		{\cal H}({\bf 15}^{s-d}  )   ^{12}_2
 		=-{\cal H}({\bf 15}^{s-d}  )   ^{13}_3
 		= -1\,,  $    $
- 2
{\cal H}({\bf 15}^b) ^{11}_1 = 
 		4 {\cal H}({\bf 15}^b) ^{12}_2
 		= 4{\cal H}({\bf 15}^b) ^{13}_3
 		=  4{\cal H}({\bf 3}_-  ) =    4{\cal H}({\bf 3}_+  )  = 
 		- 1 $
and the others are obtained by   ${\cal H}(\overline{{\bf 6}}) _{ij} = 
 	{\cal H}(\overline{{\bf 6}}) _{ji}$ 
 	and ${\cal H}({\bf 15})^{ij}_k = {\cal H}({\bf 15})^{ji}_k $.
Both the ${\bf 15}$ and ${\bf 3}$ representations contain components proportional to $\lambda_b$. The SD penguin operators transform under $SU(3)_F$ as ${\bf 3}$. Their Wilson coefficients, $C_{3 \sim 6}$, are smaller than $C_{1,2}$ by an order~\cite{Li:2012cfa}. We will first focus on the effects from the ${\bf 3}$ in the tree-level operators and then comment on those from the SD penguins at the end.

The interaction operators have been grouped according to \( SU(3)_F \), and the same grouping can be applied to the initial and final hadron states.
The antitriplet $({\bf B}_c)_i$ contains $(\Xi_c^0,\Xi_c^+,\Lambda_c^+)_i$, while ${\bf B}$ and $P$ belong to the ${\bf 8}$ representations, labelled as  
 ${\bf B}^i_j$ and $P^i_j$  including $\{ p,n,\Lambda, \Sigma ^{\pm,0} ,\Xi^{0,-}\}$ and $\{\pi^{\pm,0},\eta_8,  K^{\pm,0}, \overline{K}^0\}$ as elements, respectively~\cite{Geng:2023pkr}. 
Note that $\eta_8$ is a mixture of the physical $\eta$ and $\eta'$, which will not be discussed here.

The effective Lagrangian at the hadron level is obtained by writing   all possible $SU(3)_F$ contractions among 
$({\bf B}_c)_i, ({\bf B})^i_j$, $(P^\dagger)^i_j$ and $( {\cal H}_\pm)^{ij}_k $. 
For each contraction, we assign an unknown parameter $ \tilde{f}$ and $\tilde{g}$ for the S and P waves, respectively.
 The effective Lagrangian for S wave at the hadron level is then given by~\cite{Geng:2023pkr}
\begin{eqnarray}\label{eq22}
	{\cal L}_{ {\bf B}_c  {\bf B} P}
	&=&  (P^\dagger) ^l _n \overline{{\bf B}^k_m}  
	\left( 
	\frac{\lambda_s - \lambda_d} {2} 
	(F^{s-d}  )^{mn}_{ijkl} \epsilon^{ijn}  ({\bf B}_c )_n 
	+ \lambda_b (F^{b}  )^{imn}_{kl} 
	({\bf B}_c) _i 
	\right) 
	+(\textrm{H.c.})
	\,, \nonumber\\
	(F^{s-d}  )^{mn}_{ijkl}
	&=&
	\tilde{f}^b {\cal H}(\overline{{\bf 6}}  )_{il} \delta ^n _k  \delta ^m _ j +\tilde{f}^c 
	{\cal H}(\overline{{\bf 6}}   )_{ik}
	\delta ^n _j  (\delta^\dagger) ^m_l     +  \tilde{f}^d 
	{\cal H}(\overline{{\bf 6}}  )_{kl}
	\delta ^n _j  (\delta^\dagger) ^m_i
	+\tilde{f} ^e {\cal H}({\bf 15}^{s-d})^{ mn }_l\epsilon_{ijk} /2   
	\,, \nonumber \\
	(F^{b}  )^{imn}_{kl}  &=& 
	\tilde{f} ^e   {\cal H}({\bf 15}^{b})^{ mn }_l
	\delta ^i _k 
	+ 
	\tilde{f}_{\bf 3}^b  \delta ^i _ l  
	{\cal H}({\bf 3})^m  \delta ^n _k 
	+ 
	\tilde{f}_{\bf 3}^c   
	{\cal H}({\bf 3}  )^i \delta^m_l \delta ^n _k 
	+ 
	\tilde{f}_{\bf 3}^d  
	{\cal H}({\bf 3}  )^n  \delta_l ^m \delta ^i _k     \,,
\end{eqnarray}
where 
we have recombined ${\cal H}({\bf 3}_+)$ and ${\cal H}({\bf 3}_-)$ into ${\cal H}({\bf 3})=(1,0,0)$, and  $\tilde f$ are the $SU(3)_F$ amplitudes.
Since the S  and P waves share the same flavor structure, Eq.~\eqref{eq22} holds after  the substitution of   
$(F,\tilde f) \to ( -G , \tilde g) $~\cite{Geng:2023pkr}.  The K\"orner-Patti-Woo~(KPW) theorem~\cite{Korner:1970xq-a,Korner:1970xq-b,Korner:1970xq-c} had been applied  to  eliminate several redundant terms associated with ${\cal H}({\bf 15})$.  
We will  use $\tilde{h}$ as a generic label for either $\tilde{f}$ or $\tilde{g}$, and $\tilde{h}_ {\bf 3}$ denotes any of $\tilde{h}^{b,c,d}_ {\bf 3}$

Interestingly, 
$\tilde f ^e$ appears in both $F^{s-d}$ and $F^b$.
Even if \( \tilde{h} _{\mathbf{3}}   
\)  are dropped, one can still have nonzero $A_{CP}$ and $A_{CP}^\alpha$ from
$\tilde h ^e$,  which can be   determined from CP-even data. 
In this case $A_{CP} $ and $A^ \alpha_{CP}$  are  proportional to $\lambda_b \tilde{h}^e$.
Numerical results on CP-violating quantities in this scenario are of the order of ${\cal O}(10^{-4})$ as will be shown in the next section.
 	The sizes of $A_{CP}$  are significantly smaller than the ones of $D$ meson~\cite{LHCb:2019hro}.  One wonders if enhanced CP-violating effects can appear as those showed up in $D \to \pi^+\pi^-, K^+ K^-$. If this indeed happens, it must come from $\tilde h _ {{\bf 3}}$.

\begin{figure}[!h]
	\includegraphics[width=0.24\linewidth]{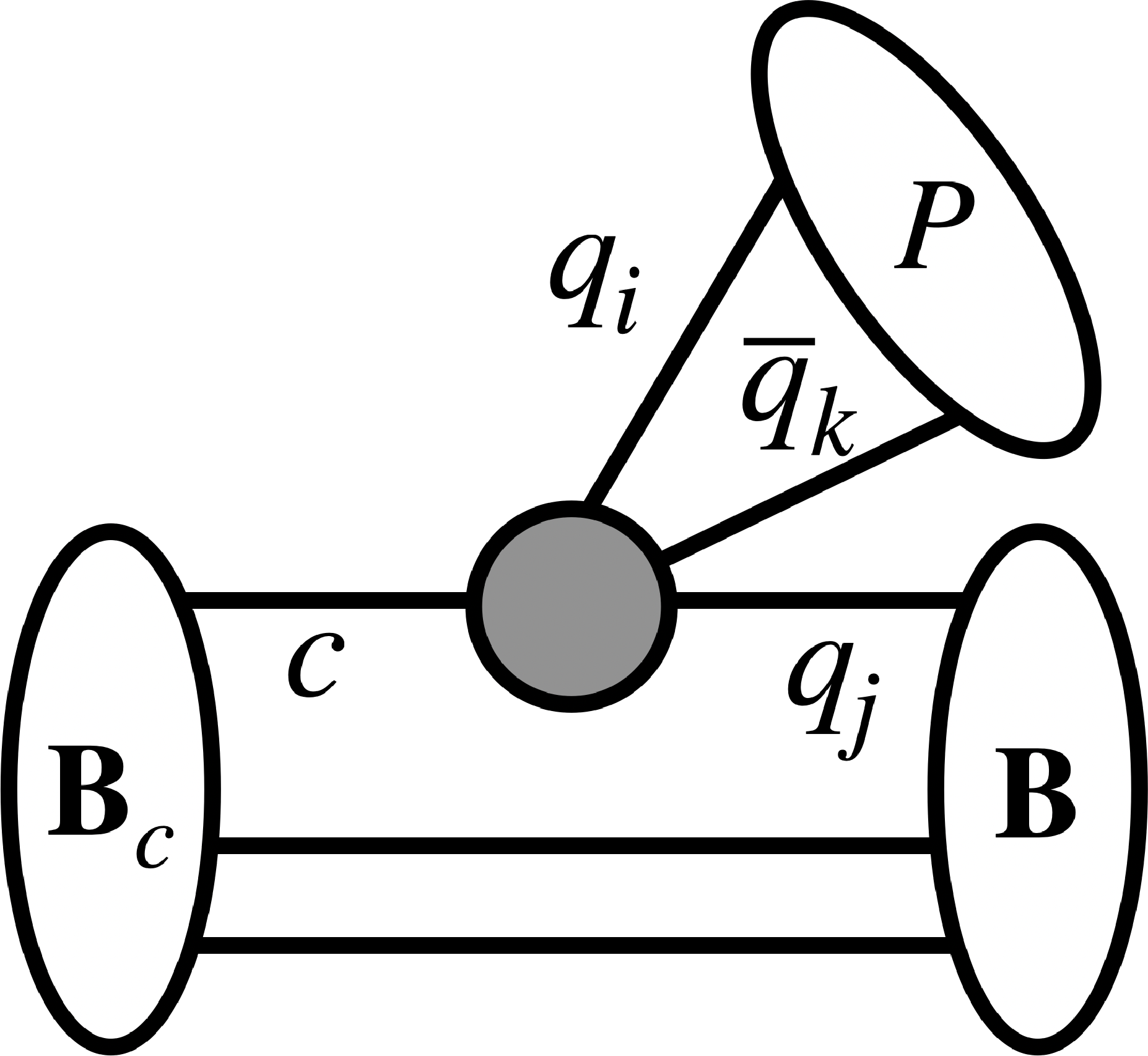}
	\caption{Topological diagram for ${\cal L}_{{\bf B}_c{\bf B}P}^{\text{Tree}}$, where the middle blob indicates the insertion of ${\cal L}_{eff}^{\text{SCS}}$.}
	\label{fig:fac}
\end{figure}

\begin{figure}[!h]
	\begin{center} 		\begin{subfigure}{0.52 \linewidth}
			\includegraphics[width=0.6\linewidth]{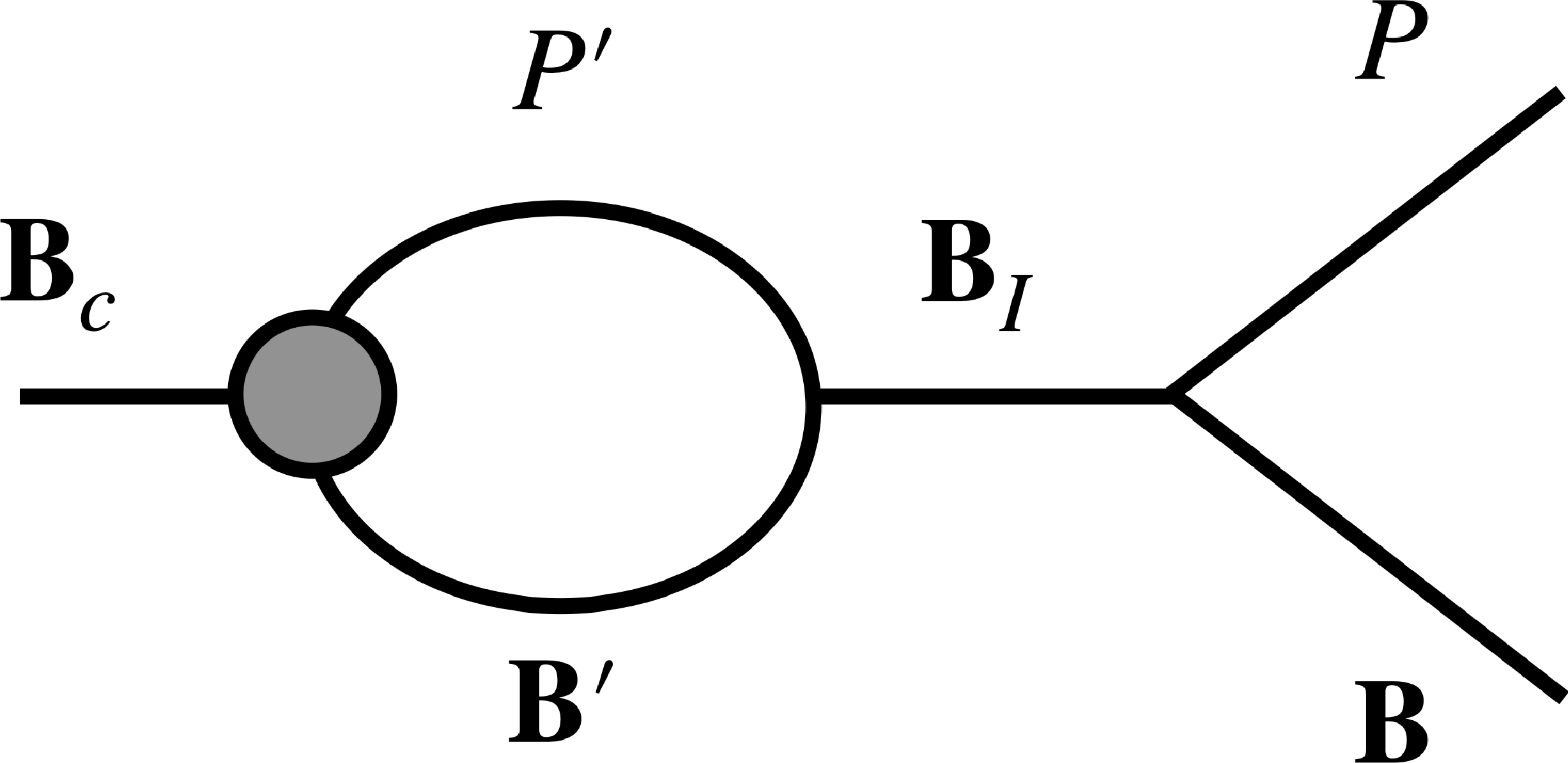}
			\caption{} \label{fig:FSRs}
		\end{subfigure}~~
		\begin{subfigure}{0.36 \linewidth}
			\includegraphics[width=0.6 \linewidth]{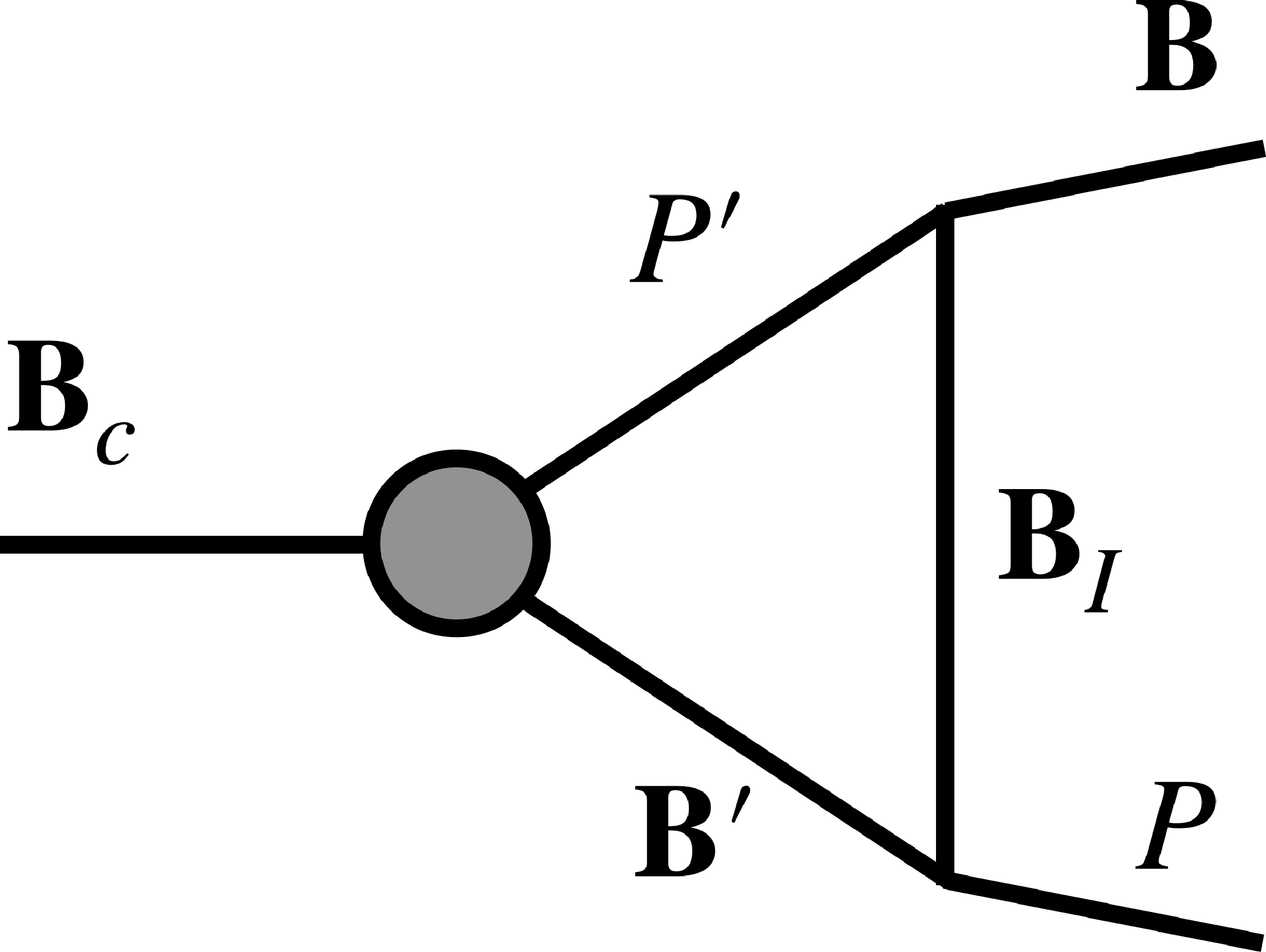}
			\caption{} \label{fig:FSRt}
		\end{subfigure}
		\caption{
			The left(right) diagram represents  the $s(t)$-channel FSR  with 
			the blob representing the SD weak transition induced by $ {\cal L} ^{\text{Tree}}  _{{\bf B} _c {\bf B} P }$. 
 	Here we consider elastic re-scattering such that ${\bf B}’$ and $P’$ belong to the same $SU(3)_F$ octets as ${\bf B}$ and $P$, respectively. }  
		\label{fig:TP}
	\end{center}
\end{figure}

   \begin{figure}[!tbh]
	\includegraphics[width=0.6\linewidth]{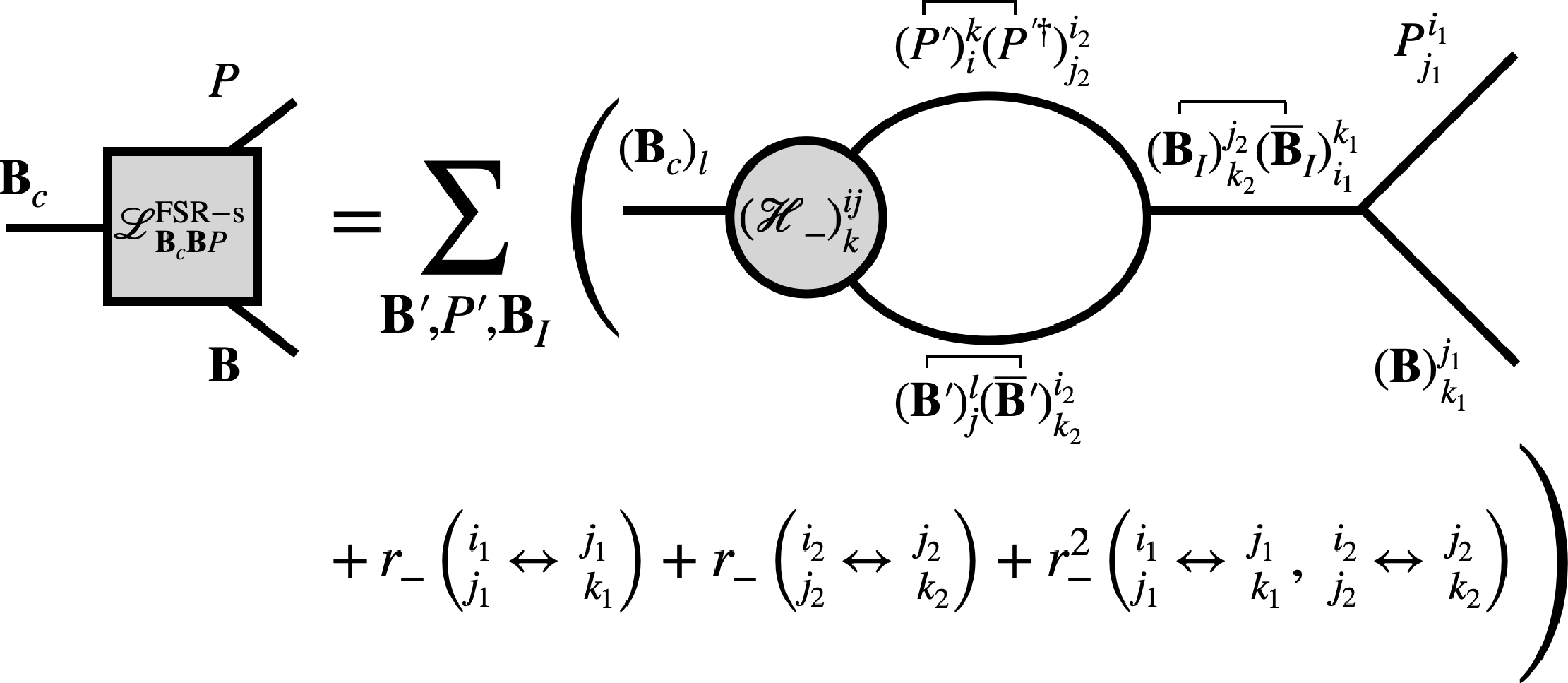}
	\caption{ \justifying
		A pictorial explanation of Eq.~\eqref{impo}. The square blob on the left-hand side represents the insertion of ${\cal L}_{{\bf B}_c {\bf B}P}^{\text{FSR-s}}$, while the circle blob on the right stands for ${\cal L}_{{\bf B}_c {\bf B}' P'}^{\mathrm{Tree}}$. The other vertices are strong couplings.
	}
	\label{fig:sum}
\end{figure}

To estimate the sizes of   $   \tilde{h}_{\mathbf{3}}$, we adopt a model analysis combining the SD tree-emissions  and  FSR  effects. Within this framework, the full effective Lagrangian for ${\bf B}_c \to {\bf B} P$ at the hadron level is decomposed into three parts:
\begin{equation}\label{RES}
	{\cal L}_{{\bf B}_c {\bf B} P}
	=  
	{\cal L}_{{\bf B}_c {\bf B} P}^{\text{Tree}}
	+ 
	{\cal L}_{{\bf B}_c {\bf B} P}^{\text{FSR-s}}
	+
	{\cal L}_{{\bf B}_c {\bf B} P}^{\text{FSR-t}}\,.
\end{equation}
The first term is induced by the SD  tree emission depicted in Fig.~\ref{fig:fac}, while the second (third) term is responsible for the LD  $s(t)$-channel 
 elastic 
re-scattering depicted in Fig.~\ref{fig:TP}, where ${\bf B}'$ and $P'$ also belong to the same $SU(3)_F$ octets as ${\bf B}$ and $ P $.
  Studies have shown that   multi-meson 
  re-scattering is suppressed compared to two-to-two particle re-scattering at the charm energy scale~\cite{Garrote:2022uub}.  We will also work with this approximation for the baryon-meson re-scattering.

The main goal is to determine the $SU(3)_F$ structure of the FSR; in other words, the relationship between $F^{s-d}$ and $F^b$ in Eq.~\eqref{eq22}. Since the former has already been determined in the global $SU(3)_F$ fit from CP-even data, we can immediately determine the latter once their relationship has been established.
   
According to  Fig.~\ref{fig:fac},  the tree-emission  diagram   has the flavor structure of 
\begin{eqnarray} \label{FAC}
	{\cal L}^{\text{Tree}}_{{\bf B}_c {\bf B} P} &=&
	(P^{\dagger})^k_i (\overline{\bf B})^l_j
	\Big(
	\tilde{F}_V^+ ({\cal H}_+)^{ij}_k
	+ \tilde{F}_V^- ({\cal H}_-)^{ij}_k
	\Big)
	({\bf B}_c)_l . 
\end{eqnarray}
The parameters $\tilde{F}_V^+$ and $\tilde{F}_V^-$ represent the overall unknowns.  By replacing $\tilde{F}_V^\pm$ with $\tilde{G}_A^\pm$ and inserting $\gamma_5$, we obtain the result for the P wave. 
We follow the   the generalized factorization approach~\cite{Buras:1985xv,Bauer:1986bm},   which introduces effective color numbers in the  S- and P-wave amplitudes as unknown parameters. 
In the most general case, the effective color numbers can be complex, and the S- and P-wave amplitudes may have different values. To account for this, we factor out these unknowns and   absorb them into the definitions of $\tilde F_V^\pm$ and $\tilde G_A^\pm$, which are treated as free parameters in our analysis.

By decomposing ${\cal H}_\pm$ according to Eqs.~\eqref{Op} and \eqref{Om}, we see 
 ${\bf 15}$, $\overline{\bf 6}$, and ${\bf 3}$ share two parameters, $\tilde{F}_V^+$ and $\tilde{F}_V^-$ in ${\cal L}^{\text{Tree}}_{{\bf B}_c {\bf B} P}$. Hence, once the amplitudes for ${\bf 15}$ and $\overline{\bf 6}$ are determined, the amplitude for ${\bf 3}$ is fixed.
The ability to achieve this in ${\cal L}^{\mathrm{Tree}}_{{\bf B}_c{\bf B}P}$ stems from  
the fact 
that ${\cal H}_+$ fixes the ratio of contributions between ${\bf 15}$ and ${\bf 3}_+$, while ${\cal H}_-$ fixes the ratio between $\overline{{\bf 6}}$ and ${\bf 3}_-$. Later, we will show that the same can be done in the $s$- and $t$-channel FSR.

For illustration, we carry out the $SU(3)_F$ analysis of the $s$-channel.
 The $s(t)$-channel
FSR 
diagram is given in Fig.~\ref{fig:FSRs}(\ref{fig:FSRt}) where the blob represents the SD weak transition induced by ${\cal L} ^{\text{Tree}}  _{{\bf B} _c {\bf B} P }$. 
The
other vertices are 
 strong couplings among  the hadrons   induced by 
$ 	{\cal L}_{{\bf B} _\pm   {\bf B}    P} ^{\text{strong}}: $
\begin{eqnarray} \label{12}
&&{\cal L}_{{\bf B}  _- {\bf B}     P} ^{\text{strong}} 
= 
g_-
\left( 
(P^\dagger) ^i_j 
(\overline{ {\bf B}} '  )  ^j _k 
+
r_-
(P^\dagger) ^j_k 
(\overline{ {\bf B}} ' ) ^i _j 
\right) ({\bf B} _  -  ) ^k_i
+(\textrm{H.c.}) \,,  
\end{eqnarray}
where 
the subscript of $\pm$ denotes the parity of ${\bf B}_\pm$, and $g_\pm$ and $r_\pm$ are constants. The $SU(3)_F$ structure of ${\cal L}_{{\bf B}_+ {\bf B} P}^{\text{strong}}$ is the same as that of ${\cal L}_{{\bf B}_- {\bf B} P}^{\text{strong}}$ and is  obtained by replacing $(g_-, r_-)$ with $(g_+, r_+)$ and inserting $i\gamma_5$ between the baryons in Eq.~\eqref{12}.
The intermediate baryons ${\bf B}_I$ can have either parity. In addition to the case where ${\bf B}_I$ belongs to the same $SU(3)_F$ octet as ${\bf B}$, we also consider the scenario where ${\bf B}_I$ is one of the low-lying negative-parity octet baryons, such as $N(1535)$, $\Sigma(1620)$, and $\Lambda(1670)$, as they are expected to have the largest couplings.

To determine \( r_+ \), we use input from the Taylor series extrapolation~\cite{General:2003sm}, which yields \( r_+ \approx 2.47 \). For \( r_- \), we perform a $\chi^2$ fit on  the experimental two-body decay widths of 
\( N(1535) \to N\pi , N \eta \), \( \Sigma(1620) \to N\overline{K}, \Lambda \pi, \Sigma \pi  \), and \( \Lambda(1670) \to N\overline{K}, \Sigma \pi , \Lambda \eta  \)~\cite{ParticleDataGroup:2022pth}.   The best-fit solution yields \( r_- = 2.45 \pm 0.81 \). 
The values of $g_\pm$ can also be extracted using the same procedure as $r_\pm$. However, their numerical values do not impact the  results in our analysis, as  they can be absorbed into the overall unknowns.  
 To account for the running of \( r_\pm \) with respect to the energy scale, we adopt a wide range of values for \( r_+ = r_- = 2.5 \pm 0.8 \) for illustration.

Now, we derive the $SU(3)_F$ structure of the $s$-channel FSR. 
We use the S wave as an illustration.  
By viewing Fig.~\ref{fig:FSRs}, we see that ${\cal L}_{{\bf B}_c {\bf B}P} ^{\text{FSR-s}}$ consists of one ${\cal L}_{{\bf B}_c {\bf B}P} ^{\text{Tree}}$ and two ${\cal L}_{{\bf B}_- {\bf B} P } ^{\text{strong}}$. 
Contracting the hadron operators according to the topology of Fig.~\ref{fig:FSRs}, we arrive at
	\begin{eqnarray}  \label{impo}
&&- 	\langle  {\cal L}_{{\bf B}_c {\bf B}P} ^{\text{FSR-s } }  \rangle
=  
  \sum_{{\bf B}_I , {\bf B} ' , P'} 
  \left\langle 
{\cal L}_{{\bf B} _I ^-  {\bf B}    P} ^{\text{strong}}
{\cal L}_{{\bf B} _I ^-  {\bf B} '   P ' } ^{\text{strong}}
{\cal L}_{{\bf B}_c {\bf B}' P' } ^{\mathrm{Tree}} 
\right\rangle_{\text{FSR-}s} \nonumber\\ 
&&=\frac{1   }{8}
	\sum _{a,b,c=1 } ^ 8  
\Big \langle	  
g^2 _- \tilde{F}_V^- 
\wick{ 
	\big( (P^\dagger) ^{i_1}_{j_1} 
	(\overline{ {\bf B}}    )  ^{j_1}_{k_1}  
	\c3  {({\bf B}_I^-)}_a     ( \lambda  _a  ) ^{k_1}_{ i_1}   \big)  
	\big( 
	\c1 {P}_c  '(\lambda_c) ^{i_2}_{j_2} 
	 (\overline{ {\c3 {{\bf B}}}}_I^- ) _a    (\lambda_a   )  ^{j_2} _{k_2}  
	\c2 {{\bf B}}_b'  (\lambda_b)   ^{k_2}_{ i_2} \big) 
	\big( 
	\c1 {P}_c^{   \prime \dagger}( \lambda_c   ) ^ k _i  	\c2 {\overline{\bf B}} _b'  ( \lambda_b  )^l _j 
	( {\cal H} _- ) ^{ij}_k     
	({\bf B}_c )_l 
	\big)  
}\Big  \rangle   \nonumber\\
&  
& \qquad \qquad+
r_- 
\left(^{i_1} _{ j_1} \leftrightarrow \,^{j_1}_{ k_1}\right) 
+r_-\left(^{i_2} _{ j_2} \leftrightarrow \,^{j_2}_{ k_2}\right) 
+ 
r_- ^2
\left(^{i_1} _{ j_1}\leftrightarrow \,^{j_1}_{ k_1}
\,,~
^{i_2} _{ j_2} 
\leftrightarrow \,^{j_2}_{ k_2}\right) \,. 
	\end{eqnarray}
Fig.~\ref{fig:sum} provides a pictorial description of the above equation, omitting the overall couplings. In the second line, we use Eq.~\eqref{12} to expand ${\cal L}_{{\bf B}_- {\bf B} P}^{\text{strong}}$ and decompose the hadrons as
$
(h)^i_j = \frac{1}{\sqrt{2}}\sum_{a=1}^8 h_a (\lambda_a)^i_j,
$
where $(\lambda_a)^i_j$ are the Gell-Mann matrices and $h = {\bf B}', {\bf B}_I^-, P'$. This decomposition arises because these hadrons belong to the ${\bf 8}$ representation. In the second equation,
$
\left(^{i_1}_{j_1} \leftrightarrow \,^{j_1}_{k_1}\right)
$ 
denotes the term obtained by swapping $i_1$ and $j_1$ in the upper indices and $j_1$ and $k_1$ in the lower indices of the first term. 
The contributions of ${\cal H}_+$ in ${\cal L}_{{\bf B}_c {\bf B}P}^{\text{FSR-s(t)}}$ vanish due to the KPW theorem.

 In the momentum basis,     the  Wick contractions 
in Eq.~\eqref{impo}
are collectively   replaced by the hadron propagators with a loop integral:
\begin{equation}      
\tilde S^- \equiv 
\int \frac{d^4 q }{48 \pi ^4 }
\frac{
g^2 _- \tilde{F}_V^- 
}{( 
q- p_{{ \bf B}_c})^2 - m _{P'}^2 }
\frac{
	p_{{ \bf B}_c}^\mu \gamma_\mu + m_{{\bf B}_I}
}{
	p_{{ \bf B}_c}^2 - m_{{\bf B}_I}^2
}  
  \frac{
	q^\mu \gamma_\mu + m_{{ \bf B}' }
}{
	q^2 - m_{{ \bf B}' }^2 
}.
\end{equation}	
Clearly, the above integral is independent of the indices \((a,b,c)\) since the masses \( m_{{\bf B}_I,{\bf B}',P'} \) remain unchanged in the \( SU(3)_F \) limit.   Substituting  the overall factor $ \tilde S^-$ defined above, Eq.~\eqref{impo} can be rewritten as:  
 	\begin{equation} \label{122}
 \begin{aligned}
	{\cal L}_{{\bf B}_c {\bf B}P} ^{\text{FSR-s } }   
	& = 
\frac{
3 	\tilde S ^-}{8}    	\sum _{a,b,c=1 } ^ 8  
		  (P^\dagger) ^{i_1}_{j_1} 
		(\overline{ {\bf B}}    )  ^{j_1}_{k_1}  
		( \lambda  _a  ) ^{k_1}_{ i_1}      
		(\lambda_c) ^{i_2}_{j_2} 
		(\lambda_a   )  ^{j_2} _{k_2}  
		(\lambda_b)   ^{k_2}_{ i_2} 
		( \lambda_c  ) ^ k _i    ( \lambda_b   )^l _j 
		( {\cal H} _- ) ^{ij}_k     
		({\bf B}_c )_l    \\
	& +  
r_- 
\left(^{i_1} _{ j_1} \leftrightarrow \,^{j_1}_{ k_1}\right) 
+r_-\left(^{i_2} _{ j_2} \leftrightarrow \,^{j_2}_{ k_2}\right) 
+ 
r_- ^2
\left(^{i_1} _{ j_1}\leftrightarrow \,^{j_1}_{ k_1}
\,,~
^{i_2} _{ j_2} 
\leftrightarrow \,^{j_2}_{ k_2}\right) \,.
\end{aligned}
\end{equation}
 Note that $\tilde S^-$ contains unknown information regarding the momentum  dependencies of $\tilde F_V^-$ and $g_\pm$ in loop integrals, so  it    is  treated   as an independent parameter. 
We can get rid of Gell-Mann matrices by the completeness relation of  
$ 
\sum _{ a } 
(\lambda_a  ) ^i_j 
(\lambda_a)  ^{k} _{l} 
= 2(  \delta ^i _{l} \delta ^{k} _j 
-
\delta ^i _j  \delta ^{k} _{l }/3) $.
Carrying out the same reduction rules  for  $ 		{\cal L}_{{\bf B}_c {\bf B}P} ^{\text{FSR-t}}$, we match Eq.~\eqref{RES} to \eqref{eq22} and find   
	\begin{eqnarray} 
		\label{new54}
		\tilde f^b &=& 
		\tilde F_V^-  
		-(r_-+ 4)  {\tilde S ^-}  +
		\sum _{\lambda= \pm } 
		( 2 r_{\lambda }^{2}-   r_{\lambda}) \tilde T^- _\lambda 
		\,, 	
		\\
		\tilde  f^c 
		&=&  -r_- (r_-+ 4)  \tilde S ^- 
		+  \sum _{\lambda= \pm }  (r_{\lambda}^2 - 2 r_{\lambda} + 3 ) \tilde T^-   _{\lambda} 
		\,,  \nonumber \\
		\tilde f^d &=& 
		\tilde F_V ^-+\sum _{\lambda= \pm } 
		(  2 r_{\lambda}^{2} - 2 r_{\lambda} - 4 ) \tilde {T}^- _{\lambda}
		\,,  ~~~	\tilde  f^e 	= 
		\tilde F_V^+  \,, \nonumber
\end{eqnarray} 
for the CKM-leading amplitudes.
Here $\tilde  T^-_{\pm }$ 
represent  the  overall unknown constants in the $t$-channel S-wave FSR  with the subscript denoting the ${\bf B}_I$ parity.  
For the CKM-suppressed amplitudes, we have that 
\begin{eqnarray}\label{master}
	\tilde{f} ^b_{\bf 3} &=&
 (1-\frac{7r_-}{2}  )    \tilde S ^- 
	+\sum _{\lambda= \pm } ( r_{\lambda}^{2} -  5 r_{\lambda} /2 + 1) \tilde T^- _{\lambda}
	\,,  \\
	 	\tilde f^c _{\bf 3} &=&  	\frac{  (r_- +1)(7r_--2 ) }{6} 
	\tilde S ^-	
-
\sum _{\lambda= \pm } 	\frac{  r_{\lambda}^{2}  +  11 r_{\lambda}  + 1 }{6} \tilde T^- _{\lambda}
	\,, \nonumber\\
	\tilde{f} ^d_{\bf 3} &=& 
  \frac{2r_--7r^2_-}{2}  
\tilde S ^-
	+ \sum _{\lambda= \pm }  \frac{(  r_{\lambda} + 1 ) ^2}{2}  \tilde T^-_{\lambda} 
- \frac{  \tilde F^+_V 
	+ 
	2
	\tilde F^-_V   }{4}    .  \nonumber 
\end{eqnarray}
The P-wave relations can be obtained by replacing 
$(
\tilde{f}, 
\tilde F_V^\pm , \tilde S ^- , \tilde T ^-_\pm  , r_\pm )$ by  
$( 
-\tilde g ,
\tilde G_A^\pm , \tilde S^+ , \tilde T ^+ _\mp  , r_\mp  )$
with $\tilde S^+(\tilde{T}^+_\pm )$ an overall unknown  in P-wave  $s(t)$-channel.
Earlier, we have determined  
 $r_-= r_+ = 2.5\pm 0.8 $. We 
 combine the summations 
 of $\lambda$ 
 in Eqs.~\eqref{new54} and  \eqref{master} and 
 redefine $\tilde{T}^\pm = (\tilde{T}^\pm _+ + \tilde{T}^\pm _-)$.

\section{Results}

With the formula given in the previous section, we determine $\tilde h^{b,c,d,e}$ from CP-conserving data first, then use Eqs.~\eqref{new54} and  \eqref{master} to determine $\tilde h^{b,c,d}_{\bf 3}$, and finally obtain CP violation predictions.

In Table~\ref{Ftilde}, we present the extracted values of $\tilde{h}^{b,c,d,e}$ obtained using the minimal $\chi^2$ fitting method described in Ref.~\cite{Geng:2023pkr}, incorporating the latest experimental data~\cite{Lee:1957qs, Belle:2024ikp, LHCb:2024tnq, BESIII:2024mgg, BESIII:2024cbr, BESIII:2024sfz, Belle-II:2024vax}. 
We stress that all experimental data available as of March 2025 have been included in our analysis. 
Given that the current data focus on CP-even quantities and $\lambda_d \gg \lambda_b$, we omitted $\tilde{h}_{\bf 3}$ in the $SU(3)_F$ fit, but it will be included when discussing CP violation later. 
The  deviation between the current and previous fitted $\tilde{h}^{b,c,d,e}$
arises primarily from  the double $Z_2$ ambiguities inherent in the $SU(3)_F$ framework~\cite{Geng:2023pkr}  when precise measurements of the Lee–Yang parameters were unavailable. In our earlier analysis, the lack of reliable input for $\gamma$ forced us to choose a solution corresponding to a negative value of $\gamma(\Lambda_c^+ \to \Lambda \pi^+)$~\cite{BESIII:2019odb}, which has since been ruled out by the recent high-precision measurement from LHCb~\cite{LHCb:2024tnq}. As a result, the newly adopted solution, consistent with the experimentally determined positive $\gamma$, leads to a different set of $\tilde h $. Nevertheless,  the predicted  ${\cal B}$ and $\alpha$ are similar to our previous work as they  are invariant under the $Z_2$ transformations. Hence, $A_{CP}$ and $A_{CP}^\alpha$ calculated from the two sets of parameters will have similar values.

The result $\chi^2$ per degree of freedom is 2.9, indicating a certain level of discrepancy.  
It is worth pointing out that a recent lattice simulation found  
${\cal B} ( \Xi_c^0 \to \Xi^- e^+ \nu_e ) = (3.58 \pm 0.12)\%$~\cite{Farrell:2025gis},  
which is about three times larger than the experimental value  
$(1.05 \pm 0.20)\%$~\cite{ParticleDataGroup:2022pth}.  
Nevertheless, the lattice result is in good agreement with the $SU(3)_F$-based prediction  
${\cal B} (\Xi_c^0 \to \Xi^-  e^+ \nu_e ) = (4.10 \pm 0.46)\%$~\cite{He:2021qnc}.  
A likely explanation for the discrepancy is that the normalization channel of $\Xi_c^0$ was underestimated in experiments~\cite{Geng:2023pkr},  
which in turn led to underestimations of all decay branching fractions of $\Xi_c^0$~\cite{Farrell:2025gis}.
Remarkably, the reconstructed prediction  
${\cal B} ( \Xi_c^0 \to \Xi^- K^+ ) = (1.26 \pm 0.04) \times 10^{-3}$  
is also about three times larger than the experimental value  
$(3.9 \pm 1.1) \times 10^{-4}$~\cite{ParticleDataGroup:2022pth},  
mirroring the same factor of discrepancy seen in the semileptonic decay.
Upcoming experimental measurements may clarify the source of this discrepancy.
In this work, however, we focus on CP violation and therefore adhere to exact $SU(3)_F$ symmetry and the original data. A comprehensive analysis of the fitting issues   with other theoretical considerations, such as $SU(3)_F$ breaking, will be presented elsewhere.

In the framework of FSR, the unknown parameters $(\tilde{F}_ V^\pm, \tilde{S}^-, \tilde{T}^-)$ and $(\tilde{G}_ A^\pm, \tilde{S}^+, \tilde{T}^+)$ can be determined using Eq.~\eqref{new54} and the inputs $\tilde{h}^{b,c,d,e}$ obtained from the global fit. Subsequently, we obtain $\tilde h _ {{\bf 3}}$ from Eq.~\eqref{master}. We observe that their absolute values are of the same order as those of $\tilde{h }^{b,c,d,e}$, which may lead to sizable CP violation.
The predicted values of $A_{CP}$ and $A_{CP}^\alpha$, based on the omission of $\tilde{h}_ {\bf 3}$, are shown in the upper rows of Table~\ref{AppA}. On the other hand, the values of $A_{CP}$ and $A_{CP}^\alpha$ obtained by solving for $\tilde{h}_{\bf 3}$ 
within the framework of FSR 
 are listed in the lower row of the same table. The values of $A_ {CP}$ are typically enhanced to the order of $10^{-3}$, consistent with the observed enhancement of CP violation in $D$ meson decays.

	\begin{table*}[t]
	\begin{tabular}{ccccccccccc}
		\hline
		&$\tilde h ^b$ & 
		$\tilde h ^c$ &
		$\tilde h ^d$ & 
		$\tilde h ^e$ 
		& $\tilde h_{{\bf 3}}^b$ & 
		$\tilde h_{{\bf 3}}^c$ &
		$\tilde h_{{\bf 3}}^d$   \\[1pt]
		\hline
		$|\tilde  f| $ &$6.75(67 ) $& $ 2.88 ( 15)$&$1.68 ( 60)$&$1.17 ( 58)  $& 
		$0.70(11)(1)  $ & 4.37(21) (97)  & 5.12(49) (2.61)   \\[2pt]
		$\delta_f$ 
		&
		0& $ 0.35  ( 6 )$&$-0.05   (12 )$&$-2.80 (  27 )$ & 
		$ 0.90(10) (62)   $&$  -3.07( 4 ) (3)  $&$0.22 (7)  (15) $
		\\[2pt]
		$|\tilde g|$ &
		$23.47 ( 1.11 )$&$9.62 ( 44 )$&$6.14( 1.91 )$&$0.40 ( 0.63)$
		&9.09(18)(4.67) &  10.46(62)(19)  & 1.99(2.25)(4.72) \\[2pt] 
		$\delta _ g  $  
		&
		$2.49 ( 11 )$&$-1.17 (  9 )$&$2.89 ( 31 )$&$ -0.97  (6.10  )$
		&
		$-1.01(6)  (3)   $&$ -0 .69(12)  (6)  $&$ -2.07	 (75 ) (1.09)  $ \\ 
		\hline
	\end{tabular}
	\caption{\justifying
	The \( SU(3)_F \) parameters are given by \( \tilde{h} = |\tilde{h}| \exp(i\delta_h) \). The values of \( \tilde{h}^{b,c,d,e} \) are determined from a global \( SU(3)_F \) fit, while those of \( \tilde{h}_{\bf{3}} \) are obtained  in the framework of FSR. 
	In the global fit for $\Lambda_c^+$, we use the absolute experimental branching fractions, as the normalization channel $\Lambda_c^+ \to p K^- \pi^+$ is well measured with minimal uncertainties~\cite{ParticleDataGroup:2022pth}. 
	For $\Xi_c$ decay data we use the absolute branching fraction for \( \Xi_c^0 \to \Xi^- \pi^+ \) while employing relative ratios for the other \( \Xi_c^0 \) decays with   \( \Xi_c^0 \to \Xi^- \pi^+ \)  the normalized channel. 
	On the other hand, we use   absolute branching fractions for 
	\( \Xi_c^+ \) with    
	\( \Xi_c^+ \to \Xi^- \pi^+ \pi^+ \)   the normalized channel~\cite{Belle-II:2024vax,ParticleDataGroup:2022pth}, as 	the three-body decays are not parameterized in our  framework of the $SU(3)_F$ symmetry.
	The magnitudes \( |\tilde{h}| \) and phases \( \delta_h \) are expressed in units of \( 10^{-2}G_F\,\text{GeV}^2 \) and radians, respectively.
	Values within parentheses represent the last two digits of the uncertainties, such as \( 6.75(67) = 6.75 \pm 0.67 \). The first uncertainties arise from the experimental input, while the second uncertainties arise from the hadron couplings. }	\label{Ftilde}
\end{table*}

\begin{table*}[t]
	\begin{tabular}{lccc l ccc }
		\hline
		Channels &${\cal B}$&$A_{CP}  $ & $ A^\alpha_{CP}  $ 	&	Channels &${\cal B}$&$A _ {CP}  $ & $ A^\alpha_{CP}  $ \\[2pt]
		\hline
		\multirow{2}{*}{  $\Lambda_{c}^{+} \to p \pi^{0} $}
		& \multirow{2}{*}{  $ 0.18 (2 )$}&$ -0.01 ( 7 )$&$ -0.15 ( 13 )$& 
		\multirow{2}{*}{  $\Xi_{c}^{0} \to \Sigma^{+} \pi^{-} $ }& \multirow{2}{*}{  0.26(2) }&0&0\\
		&&$ 0.01 ( 15 )(45) $&$ 0.55 ( 20 )(61) $ &&&
		$ 0.71 ( 15 )(6) $&$ -1.83 ( 10 )(15) $\\[2pt]
		\multirow{2}{*}{  $\Lambda_{c}^{+} \to n \pi^{+} $}& \multirow{2}{*}{  $ 0.68 ( 6 )$ }&$ 0.0 ( 1 )$&$ 0.03 ( 2 )$&
		\multirow{2}{*}{ $\Xi_{c}^{0} \to \Sigma^{0} \pi^{0} $}& \multirow{2}{*}{  $ 0.34(3 )$}&$ -0.02 ( 4 )$&$ 0.01 ( 1 )$\\
		&&$ -0.02 ( 7 )(28) $&$ 0.30 ( 13 )(41) $ &&&
		$ 0.44 ( 24 )(17) $&$ -0.43 ( 31 )(16) $ \\[2pt]
		\multirow{2}{*}{  $\Lambda_{c}^{+} \to \Lambda  K^{+} $}& \multirow{2}{*}{  $ 0.62 ( 3 )$}&$ 0.00 ( 2 )$&$ 0.03 ( 2 )$&
		\multirow{2}{*}{   $\Xi_{c}^{0} \to \Sigma^{-} \pi^{+} $}& \multirow{2}{*}{  $ 1.76 (5  )$}&$ 0.01 ( 1 )$&$ -0.01 ( 1 )$ \\
		&&$ -0.15 ( 13 )(9) $&$ 0.50 ( 9 )(21) $&&&
		$ 0.12 ( 6 )(2) $&$ -0.22 ( 5 )(21) $\\[2pt]
		\multirow{2}{*}{  $\Xi_{c}^{+} \to \Sigma^{+} \pi^{0} $}& \multirow{2}{*}{  $2.69 ( 14 )$}& $ -0.02 ( 6 )$&$ 0.07 ( 4 )$&
		\multirow{2}{*}{  $\Xi_{c}^{0} \to \Xi^{0} K_{S/L} $ }& \multirow{2}{*}{  0.38(1) }&0&0\\
		&&$ 0.05 ( 7 )(8) $&$ -0.23 ( 3 )(15) $&&&
		$ 0.18 ( 3 )(5) $&$ -0.38 ( 2 )(11) $ \\[2pt]
		\multirow{2}{*}{ $\Xi_{c}^{+} \to \Sigma^{0} \pi^{+} $}&
		\multirow{2}{*}{ $ 3.14 ( 10 )$  }&$ 0.0 0 ( 1 )$&$ -0.02 ( 1 )$&
		\multirow{2}{*}{ $\Xi_{c}^{0} \to \Xi^{-} K^{+} $}& \multirow{2}{*}{ $ 1.26 ( 4)$}&$ 0.00 ( 1 )$&$ 0.01 ( 1 )$\\
		&&$ 0.05 ( 8 )(7) $&$ -0.24 ( 6 )(13) $  &&&
		$ -0.12 ( 5 )(2) $&$ 0.21 ( 4 )(2) $ \\[2pt]
		\multirow{2}{*}{
			$\Xi_{c}^{+} \to \Xi^{0} K^{+} $}&  \multirow{2}{*}{$ 1.30 ( 10 )$ }&$ 0.0 0 ( 0 )$&$ -0.02 ( 1 )$&
		\multirow{2}{*}{  $\Xi_{c}^{0} \to p K^{-} $ }& \multirow{2}{*}{  0.31(2)  }&0&0 \\
		&&$ 0.01 ( 6 )(17) $&$ -0.23 ( 9 )(52) $ &&&$ -0.73 ( 18 )(6) $&$ 1.74 ( 11 )(14) $ \\[2pt]
		\multirow{2}{*}{ $\Xi_{c}^{+} \to \Lambda \pi^{+} $}&
		\multirow{2}{*}{  $ 0.18 ( 3 )$}&$ -0.01 ( 2 )$&$ 0.0 ( 0 )$&
		\multirow{2}{*}{ $\Xi_{c}^{0} \to n K_{S/L} $}& \multirow{2}{*}{  0.86(3) }&0&0 \\
		&&$ -0.31 ( 21 )(13) $&$ 0.96 ( 25 )(44) $ &&&
		$ -0.14 ( 3 )(4) $&$ 0.27 ( 2 )(7) $ \\[2pt]
		\multirow{2}{*}{ $\Xi_{c}^{+} \to p K_s $}			
		&			\multirow{2}{*}{  $ 1.55 ( 7 )$ }
		&0&0&
		\multirow{2}{*}{ $\Xi_{c}^{0} \to \Lambda \pi^{0} $}
		&		\multirow{2}{*}{  $ 0.06(2  )$} &$ 0.02 ( 3 )$&$ 0.0 ( 1 )$\\
		&&$ -0.13 ( 3 )(4) $&$ 0.22 ( 3 )(7) $&&&$ -0.12 ( 18 )(10) $&$ 0.69 ( 8 ) ( 43)$
		\\[2pt]
		\hline	
	\end{tabular}
	\caption{\justifying
	The ${\cal B}$, $A_{CP}$, and $A_{CP}^\alpha$ are presented in units of $10^{-3}$. The upper rows of $A_{CP}$ and $A_{CP}^\alpha$ are obtained by taking $\tilde h _{{\bf 3}}  = 0$, while the lower rows are obtained through the FSR mechanism. The uncertainties are quoted in the same manner as Table~\ref{Ftilde}. 
	There may be some $SU(3)_F$  breaking effects that affect the results on the order of $10^{-1}$, as found in the hyperons. 
}
\label{AppA}
\end{table*}

\section{Discussion  and conclusion }
Under the U-spin symmetry, an $SU(2)$ subgroup of $SU(3)_F$ between $d$ and $s$, exchanging $d$ and $s$ in the initial and final states  can be   viewed as taking
the transformation  of 
$
(
\lambda_s - \lambda_d  ,
\lambda_b  
)  
\rightarrow 
(
-	\lambda_s +  \lambda_d   , 
\lambda_b 
)  \,.$
Since CP  asymmetries are  proportional to 
Im$(\lambda_b ^* (\lambda_s-\lambda_d))$, 
they flip signs after exchanging $d$ and $s$ in the initial and final states. 
This kind of $SU(3)_F$ relation  provides useful information to test the  CKM mechanism  model-independently~\cite{Deshpande:1994ii}.
From the U-spin symmetry we have~\cite{Wang:2019dls,He:2018joe} 
\begin{eqnarray}\label{uspi}
	&&	A_{CP}   ( 
	\Xi_c^0 \to \Xi^0 K_{S/L}^0 
	)  =- 
	A_{CP}   ( 
	\Xi_c^0 \to n K_{S/L}^0 
	) \;,\nonumber\\
	&& A_{CP}   ( 
	\Xi_c^0 \to \Sigma^- \pi^+ 
	)  =- 
	A_{CP}   ( 
	\Xi_c^0 \to \Xi^- K^+  
	)  \,, \nonumber\\
	&& A_{CP}   ( 
	\Xi_c^0 \to \Sigma^+ \pi^- 
	)  =- 
	A_{CP}   ( 
	\Xi_c^0 \to p K^-
	)     \,.
\end{eqnarray}
These direct relations also apply when substituting \( A^\alpha_{CP} \)  for \( A_{CP}  \). 

The above U-spin relations  may   help reduce systematic errors in experiments, in the same spirit as $A_{CP}(D^0 \to \pi^+\pi^-)-A_{CP}(D^0 \to K^+K^-)$.
Specifically, we recommend measuring \( A_{CP} (\Xi_c^0 \to p K^-) - A_{CP} (\Xi_c^0 \to \Sigma^+ \pi^-) \) and \( A_{CP}^\alpha (\Xi_c^0 \to p K^-) - A_{CP}^\alpha (\Xi_c^0 \to \Sigma^+ \pi^-) \), which, according to our analysis, are found to be \( (-1.44 \pm 0.35) \times 10^{-3} \) and \( (3.57  \pm 0.36) \times 10^{-3} \), respectively. All the final states are charged, making them   ideal channels to probe CP violation in baryon decays.
The intuitive reason why these two channels exhibit large CP violation can be traced to   two similar topological diagrams, as shown in the graphic abstract. The diagrams are comparable in magnitude,  allowing for significant interference between $V_{cd}^ * V_{ud}$ and $V_{cs}^ *  V_{us}$.

  Data from LHCb revealed significant U-spin breaking in \(D^0\) decays~\cite{LHCb:2022lry}. Resonances \(f(1710)\) and \(f(1790)\), which are close in mass to \(D\) mesons, are thought to contribute significantly to \(SU(3)_F\) breaking in \(D^0\) decays~\cite{Cheng:2010ry,Schacht:2021jaz}. Conversely, no similar mechanism is expected in charmed baryons, and good \(SU(3)_F\) fits   indicate that $SU(3)_F$ is applicable to the precision we are interested in~\cite{Geng:2023pkr,Farrell:2025gis,He:2021qnc}.
  Large deviations of U-spin symmetry  are not expected,   providing valuable experimental tests for the SM. 
  To be conservative, one should note that $SU(3)_F$ breaking may affect our predictions on the order of $10^{-1}$ as found in the hyperons. Nevertheless, the breaking should not affect the overall sizes of CP asymmetries,  found to be $10^{-3}$ in this work. 

Before ending the discussion, we would like to comment on the SD penguin operators which have been neglected so far. 
We estimate their sizes with the naive factorization.
In this framework, the SD penguin operators will scale  the last term $-(\tilde{F}_V^+ + 2 \tilde{F}_V^- )/4$ in 
$\tilde f _{{\bf 3}}^d$  in Eq.~\eqref{master}
 by a factor of $( 1 + {\cal R}_+) $ with
 ${\cal R}_\pm = (3 C_4 + C_3) / C_1 
 \pm   m_K^2 (6 C_6 +2  C_5)/ (m_s m_c C_1) $ 
and similarly  $-(\tilde{G}_A^+ + 2 \tilde{G}_A^-)/4 $ in $\tilde{g}_{{\bf 3}}^d$ by a factor  of $(1 + {\cal R}_-) $. We have checked that the corrections on $A_{CP}$ and $A_{CP}^\alpha$ are less than $10\%$, which can be neglected. 

It should also be noted that although our framework is built upon the SM, our $SU(3)_F$ fit may partially absorb new physics contributions. From the results we have obtained, we would like to conclude that there is no clear  evidence indicating the need for physics beyond the SM to explain the data.
A significant number of charmed baryons will be produced in the near future. 
Thus far, Belle’s dataset has achieved measurements of \(A_{CP}\) and \(A_{CP}^\alpha\) with precision at the percent level~\cite{Belle:2022uod}. 
Belle II is expected to increase production, achieving precision  at  the subpercent level. 
At LHCb, the production rates of charmed baryons are suppressed to around 30\% compared to \(D\) mesons~\cite{LHCb:2018weo}, but the precision of \(A_{CP}\) may also reach \(10^{-3}\). Following the High Luminosity upgrade~\cite{Apollinari:2017lan}, the integrated luminosity is expected to increase tenfold.
In the future, the Super \(\tau\)-Charm Facility may also provide valuable data on these decays~\cite{Achasov:2023gey, Cheng:2022tog}. The enhancement identified in this work facilitates the first observation of CP violation in two-body charmed baryon decays.

\section*{Conflict of interest}
The authors declare that they have no conflict of interest.

\begin{acknowledgments}
This work is supported in part by the National Key Research and Development Program of China 2020YFC2201501,  the 
 National Natural Science Foundation of  China (12090064, 12205063, 12375088, and W2441004), and  the China Postdoctoral Science Foundation  2023M742293.
\end{acknowledgments}

\section*{Author contributions}
The project originated from a mutual discussion on possible large CP violation in charmed baryon decays. Xiao-Gang He formulated the overall strategy, and Chia-Wei Liu developed key methods to pursue the analysis. Both authors contributed equally to the research and manuscript preparation.

\end{document}